\documentclass[12pt]{article}

\usepackage{amsmath,amssymb,amsfonts,latexsym}
\usepackage{rotating}

\newcommand{\Ha}{H^{(\alpha,\beta)}}
\newcommand{\tHa}{\tilde{H}^{(\alpha,\beta)}_{+}}
\newcommand{\ttHa}{\tilde{H}^{(\alpha,\beta)}_-}

\newcommand{\tha}{\tilde{h}^{(\alpha,\beta)}_{+}}
\newcommand{\ttha}{\tilde{h}^{(\alpha,\beta)}_{-}}
\newcommand{\bha}{\bar{h}^{(\alpha,\beta)}}

\newcommand{\ro}{\rho_{(\alpha,\beta)}}
\begin{document}
{
\begin{center}
 \Large\bf {A generalized Swanson Hamiltonian in a second-derivative pseudo-supersymmetric framework }
 \end{center}

 \begin{center}
Bijan Bagchi\footnote{bbagchi123@gmail.com},
Abhijit Banerjee\footnote{abhijit.banerjee.81@gmail.com, abhijit\_banerjee@hotmail.com},
 Partha Mandal\footnote{parthamandal1999@gmail.com}
 \end{center}

 \begin{center}
 $^{1,3}$ Department of Applied Mathematics, University of Calcutta, 92 Acharya Prafulla Chandra Road, Kolkata, India-700009\\
 $^{2}$ Department of Mathematics, Krishnath College, Berhampore, Murshidabad, West Bengal India-742101\\
 \end{center}

\begin{center}
\large{\bf{Abstract}}
\end{center}
We study a generalized scheme of Swanson Hamiltonian from a second-derivative pseudo-supersymmetric approach. We discuss plausible choices of the underlying quasi-Hamiltonian and consider the viability of applications to systems like the isotonic oscillator and CPRS potential.\\

Keywords: PT-symmetry, Pseudo-Hermiticity, Swanson Hamiltonian, pseudo-supersymmetry.\\

PACS Nos: {03.65.;02.30.}, MSC {81Q12; 81Q60}
\newpage
\section{\label{intro}Introduction:}
\numberwithin{equation}{section}

Following Bender and Boettcher's conjecture in 1998 \cite{Ben1} that a class of  quantum Hamiltonians invariant under the combined action of parity ($\mathcal{P}$) and time($\mathcal{T}$) can possess a real bound-state spectrum except when the symmetry is spontaneously
broken, in which case their complex eigenvalues should come in conjugate pairs, there has been a growing interest in the study of such systems \cite{Ben2,Mos1}. That the mathematical foundation of $\mathcal{PT}$-symmetry has its roots in the theory of pseudo-Hermitian operators was subsequently shown by Mostafazadeh in a series of papers \cite{Mos2, Mos3,Mos4}. For the reality of the spectrum the Hamiltonian $H$ is needed to be Hermitian with respect to a positive-definite inner product $<.,.>_+$  on the Hilbert space $\mathbb{H}$ in which $H$ is acting. This inner product can be expressible in terms of a metric-induced defining inner product as \cite{Mos2}
\begin{equation}
  <.,.>_+  =<.,\zeta.>
\end{equation}
       where the positive-definite metric operator $\zeta:\mathbb{H}\rightarrow \mathbb{H}$ belongs to the set of all Hermitian invertible operators. The Hilbert space $\mathbb{H}$ equipped with the above inner product is identified as the physical Hilbert space $\mathbb{H}_{\mbox{phys}}$. The pseudo-Hermiticity of $H$ is given by
\begin{equation}\label{psu-1}
H^\dag=\zeta H \zeta^{-1}
\end{equation}
that serves as one of the plausible necessary and sufficient conditions for the reality of spectrum. It may be mentioned here that in a recent work it has also been pointed out \cite{Fer} that in spite of the manifest non-Hermiticity of the Hamiltonian, unitarity of the time evolution of the system is achieved in a properly amended physical Hilbert space.

An observable $\mathcal{O}\in \mathbb{H}_{\mbox{phys }}$ is related to the Hermitian operator $o\in \mathbb{H}$ by means of a similarity transformation $\mathcal{O}=\rho^{-1}o\rho$ where $\rho$ is unitary and  $\zeta$  is furnished in the factorized form $\zeta=\mathcal{O}^{\dag}\mathcal{O}$.
 Further  $\rho$ is given by
\begin{equation}\label{psu-3}
\rho=\sqrt{\zeta}:~\mathbb{H}_{\mbox{phys}}\rightarrow  \mathbb{H}
\end{equation}
Given a knowledge of $\rho$, the equivalent Hermitian Hamiltonian $h$ may be identified as
\begin{equation}\label{psu-4}
h=\rho H \rho^{-1}.
\end{equation}

  Swanson \cite{Swan} considered a specific type of pseudo-Hermitian quadratic Hamiltonian connected to an extended harmonic oscillator problem. He proposed a general form namely,
\begin{equation}\label{SW-H}
\Ha=\omega \eta^\dag \eta +\alpha \eta^2 +\beta {\eta^\dag}^2 +\frac{1}{2}\omega
\end{equation}
 in terms of the usual annihilation and creation operators $\eta$ and $\eta^\dag$ of the harmonic oscillator obeying the canonical commutation relation $\left[\eta,\eta^\dag\right]=1$. In (\ref{SW-H}), $\omega, \alpha$ and $\beta$ are real constants. It is evident that with $\alpha\neq\beta$, the Hermitian character of $\Ha$ is lost. Nonetheless, it is $\mathcal{PT}$-symmetric and as typical with such models, support a purely real, positive spectrum over a certain range of parameters. Swanson Hamiltonian as a toy model has been variously used to investigate non-Hermitian systems in different representations. These include, to name a few, exploring the choice of a unique and physical metric operator to set up an equivalent Hermitian system \cite{Jones,Mus}, identifying the relevant group structure of the Hamiltonian \cite{Qes1,Asi}, seeking quasi-Hermitian \cite{Qes2} and pseudo-supersymmetric (SUSY) extensions \cite{Yes1}, looking for $\mathcal{N}$-fold SUSY connections \cite{BT}, writing down minimum length uncertainty relations resulting from non-commutative algebras \cite{Fri1, Fri2}, investigating a relevant R-deformed algebra \cite{Rr}, deriving supercoherent states \cite{CDMT} and more recently, studying classical and quantum dynamics for it \cite{Gra}.

The hidden symmetry structure of Swanson Hamiltonian and of its Hermitian equivalence has also been pursued from the point of view of a generalized quantum condition $\left[\eta,\eta^\dag\right]\neq 1$ using the representation \cite{bag1}

\begin{equation}\label{eta-0}
\eta=a(x)\frac{d}{dx}+b(x) \qquad a(x),b(x)\in \mathbb{R}
\end{equation}
 for which
 \begin{equation}\label{gen-qnt-cond}
 \left[\eta,\eta^\dag\right]=2ab'-aa'',
 \end{equation}
where $'=\frac{d}{dx}$. This enables us to connect a large class of physical systems for suitable choices of the functions a and b.
A generalized quantum condition has been found to have relevance to position dependent mass (PDM) systems \cite{bag1}. In a recent analysis \cite{yes} a particular class of the generalized quantum condition was also analyzed in the context of a generalized $\eta$ given by (\ref{eta-0}).

 In the following we study the non-Hermitian Swanson Hamiltonian from a second-derivative SUSY (SSUSY) perspective \cite{And1,And2,And3,Fer1,Fer2,bag2} by resorting to second-derivative representations of the factorization operators.  In the literature two-component SSUSY schemes have found applications to non-trivial quantum mechanical problems such as those of coupled channel problems \cite{Can} and transparent matrix potentials \cite{Pup}. The representative character of SSUSY is controlled by a quasi-Hamiltonian $K$ which is a fourth-order differential operator (in other words, $K$ is second order in the  Schr\"odinger operator). This leads to the picture of the so-called ``polynomial SUSY".

 The plan of the paper is as follows:\\
 In section \ref{ssyehh} we discuss the SSUSY realization of the equivalent Hermitian Hamiltonian representation of $\Ha$; in section \ref{ssyehh1} we elaborate upon the corresponding pseudo-SUSY aspects and construct a pseudo-superalgebra in terms of pseudo-supercharges. Here we also write down the underlying form of the quasi-Hamiltonian $\tilde{K}$ ; in section 4 we construct some classifications of its Hermitian equivalent counterpart $K$; in sections 5 and 6 we respectively address the specific examples of the isotonic oscillator and the CPRS potential as possible applications of our scheme; finally in section 7 we give a conclusion of our work.

\section{\label{ssyehh}SSUSY realization of equivalent Hermitian Hamiltonian:}
\numberwithin{equation}{section}

For the generalized representation of $\eta$   as in (\ref{eta-0}), the Swanson Hamiltonian (\ref{SW-H}) can be cast in the explicit form \cite{bag1}
\begin{equation}\label{SW-H-expl}
\Ha\rightarrow\tHa=-\frac{d}{dx}\tilde{a}^2(x)\frac{d}{dx}+\tilde{b}(x)\frac{d}{dx}+\tilde{c}(x)
\end{equation}
where
\begin{eqnarray}
\tilde{a}&=&\sqrt{\tilde{\omega}}a,\quad \tilde{\omega}=\omega-\alpha-\beta >0\nonumber\\
\tilde{b}&=&-(\alpha+\beta)aa'+2\alpha ab-2\beta a(b-a'),\nonumber\\
\tilde{c}&=&-\omega (ab)'+(\alpha+\omega)b^2+\alpha ab'-\beta a(b-a')'+\beta (b-a')^2 +\frac{\omega}{2}.
\end{eqnarray}
 Note that $\tilde{H}^{(\alpha,\beta)}_{+}$ is taken to represent the seed Hamiltonian of a two-component SSUSY family denoted by a $(+)$-suffix.

The first-derivative term of (\ref{SW-H-expl}) can be removed by employing the transformation
\begin{equation}\label{rho}
\ro=e^{-\frac{1}{2}\int^x \frac{\tilde{b}}{\tilde{a}^2}dx}
\end{equation}
yielding an equivalent Hermitian Hamiltonian $\tha$ of (\ref{SW-H-expl}):
\begin{equation}\label{eqHH}
\tha=\ro \tHa \ro^{-1}
\end{equation}
Here $\tha$ reads explicitly
\begin{equation}\label{h-0}
\tha= -\frac{d}{dx}\tilde{a}^2(x)\frac{d}{dx}+\tilde{V}^{(\alpha,\beta)}_{+}(x)
\end{equation}
while $\tilde{V}^{(\alpha,\beta)}_{+}(x)$ is given by
\begin{eqnarray}
\tilde{V}^{(\alpha,\beta)}_{+}(x) &=& \left[\frac{(\alpha-\beta)^2}{\tilde{\omega}}+\tilde{\omega}+2(\alpha+\beta)\right]b\left(b-\frac{\tilde{a}'}{\sqrt{\tilde{\omega}}}\right)
        -\left[\tilde{\omega}+(\alpha+\beta)\right]\frac{\tilde{a}b'}{\sqrt{\tilde{\omega}}}\nonumber\\
    &+&\frac{\alpha+\beta}{2\tilde{\omega}}\tilde{a}\tilde{a}''+\frac{1}{4\tilde{\omega}}\left[\frac{(\alpha-\beta)^2}{\tilde{\omega}}+2(\alpha+\beta)\right]\tilde{a}'^2
       +\frac{\tilde{\omega}+\alpha+\beta}{2}.\label{V-0}
\end{eqnarray}

 To proceed with the SSUSY construction we propose the existence of a partner Hamiltonian to $\tilde{h}^{(\alpha,\beta)}_{+}$ namely, the $\tilde{h}^{(\alpha,\beta)}_{-}$  operator such that they together form the diagonal entries of a $2 \times 2$ matrix as given by
\begin{equation}\label{H-s}
\mathfrak{H}=\left(
    \begin{array}{cc}
      \tha & 0 \\
      0 & \ttha \\
    \end{array}
  \right)
  \end{equation}
  where
  \begin{equation}\label{h-s}
  \tha=-\frac{d}{dx}\tilde{a}^2(x)\frac{d}{dx}+\tilde{V}^{(\alpha,\beta)}_{+}(x),
  \quad\ttha=-\frac{d}{dx}\tilde{a}^2(x)\frac{d}{dx}+\tilde{V}^{(\alpha,\beta)}_{-}(x),
  \end{equation}
 and $\tilde{V}^{(\alpha,\beta)}_{-}(x)$ be the potential of $\ttha$.\\

  Coresponding to $\mathfrak{H}$, the associated supercharges of the underlying SSUSY theory can be projected in terms of second order differential operators $\mathcal{A}^\pm$
  \begin{equation}\label{sup-ch}
  Q^+ = \left(
          \begin{array}{cc}
            0 & 0 \\
            \mathcal{A}^- & 0 \\
          \end{array}
        \right),\quad Q^- = \left(
          \begin{array}{cc}
            0 & \mathcal{A}^+ \\
             0& 0 \\
          \end{array}
        \right)
  \end{equation}
  It is easy to see that $\mathfrak{H}$ commutes with both $Q^{+}$ and $Q^{-}$
  \begin{equation}
    [Q^{+},\mathfrak{H}]=0,\quad [Q^{-},\mathfrak{H}]=0
  \end{equation}
  provided the following intertwining relations hold
  \begin{equation}\label{susy-int}
  \mathcal{A}^- \tha=\ttha \mathcal{A}^-,\quad \mathcal{A}^+ \ttha=\tha \mathcal{A}^+.
  \end{equation}
  $Q^{+}$ and $Q^{-}$ may be combined to get the quasi-Hamiltonian $K$ which is quadratic in $\mathfrak{H}$
  \begin{equation}\label{quasi-h}
  K = Q^+ Q^- + Q^- Q^+ = \mathfrak{H}^2 - 2 \lambda \mathfrak{H} +\mu, \qquad \lambda,\mu \in \mathbb{R}.
  \end{equation}
The passage of SUSY to SSUSY is thus a transition from $H^{\alpha,\beta}\rightarrow K$.

\section{\label{ssyehh1}Pseudo-SUSY :}

Note that the similarity transformations (\ref{psu-3}) actually give the equivalent relationships
\begin{eqnarray}
\tHa &=& \ro^{-1}\tha\ro, \label{n-psu-1}\\
\ttHa &=& \ro^{-1}\ttha\ro \label{n-psu-2}
\end{eqnarray}
 which provide the expected connection between $\tha$ and $\ttha$ with their non-Hermitian counterparts $\tHa$ and $\ttHa$.

 It is interesting to see that if we right-multiply the first of the above relation by $\ro^{-1}\mathcal{A}^+ \ro$ then we at once deduce using (\ref{susy-int})
\begin{eqnarray*}
\tHa \ro^{-1}\mathcal{A}^+ \ro &=& (\ro^{-1}\tha\ro)(\ro^{-1}\mathcal{A}^+ \ro)\nonumber\\
&=& \ro^{-1}(\tha \mathcal{A}^+)\ro \nonumber\\
&=&\ro^{-1}(\mathcal{A}^+ \ttha)\ro \nonumber\\
&=& (\ro^{-1}\mathcal{A}^+ \ro)(\ro^{-1}\ttha\ro)\nonumber\\
&=&(\ro^{-1}\mathcal{A}^+ \ro)\ttHa \nonumber\\
\end{eqnarray*}
\begin{equation}
  \mbox{i.e.}~~ \tHa \vartheta^+ =\vartheta^+ \ttHa, \label{psusy-int-1}
\end{equation}
where $\vartheta^+=\ro^{-1}\mathcal{A}^+ \ro$.

On the other hand, right-multiplying (\ref{n-psu-2}) by $\ro^{-1}\mathcal{A}^- \ro$ we obtain
\begin{equation}\label{psusy-int-2}
\ttHa \vartheta^- =\vartheta^- \tHa
\end{equation}
where $\vartheta^-=\ro^{-1}\mathcal{A}^- \ro$.

The operators $\vartheta^+,\vartheta^-$ are thus pseudo-adjoint in the sense that \cite{Mos2,CDMT, DG}
\begin{equation}
{(\vartheta^-)}^\sharp =\zeta^{-1}{(\vartheta^-)}^\dag \zeta =\ro^{-2}(\ro^{-1}\mathcal{A}^- \ro)^\dag\ro^2=\ro^{-1}\mathcal{A}^+ \ro= \vartheta^+.
\end{equation}
With respect to these pseudo-adjoint operators $\vartheta^\pm$, we therefore conclude that the equations (\ref{psusy-int-1}) and (\ref{psusy-int-2}) play the role of intertwining relations between the non-Hermitian Hamiltonians $\tHa$ and $\ttHa$. We are thus in a position to construct a pseudo-superalgebra in terms of the pseudo-supercharges

\begin{equation}
   Q=\left(
                                                                                                                                  \begin{array}{cc}
                                                                                                                                    0 & 0 \\
                                                                                                                                    \vartheta^- & 0 \\
                                                                                                                                  \end{array}
                                                                                                                                \right)
,\quad Q^\sharp=\left(
                     \begin{array}{cc}
                       0 & \vartheta^+ \\
                       0 & 0 \\
                     \end{array}
                   \right)
.
\end{equation}

The underlying quasi-Hamiltonian $\tilde{K}$ for such a pseudo-SUSY scenario reads
\begin{equation}\label{psu-qs-ham-1}
\tilde{K}=QQ^\sharp + Q^\sharp Q =\left(
                                                              \begin{array}{cc}
                                                                \vartheta^+ \vartheta^- & 0 \\
                                                                0 & \vartheta^- \vartheta^+ \\
                                                              \end{array}
                                                            \right)
                                                            \end{equation}
It is quadratic in
\begin{equation}\label{5.8}
\tilde{\mathcal{H}}=\left(
                                     \begin{array}{cc}
                                       \tHa & 0 \\
                                       0 & \ttHa \\
                                     \end{array}
                                   \right)
\end{equation}
since it is expressible in the form
\begin{equation}\label{psu-qs-ham-2}
\tilde{K}=\tilde{\mathcal{H}}^2 -2\lambda \tilde{\mathcal{H}}+\mu;\quad \lambda,\mu \in \mathbb{R}.
\end{equation}
Evidently $\left[\tilde{\mathcal{H}},Q\right]=\left[\tilde{\mathcal{H}},Q^\sharp\right]=0$. The latter along with $(Q)^{2}=(Q^\sharp)^{2}=0$ and (\ref{psu-qs-ham-1}) define the algebra of $N=2$ pseudo-SUSYQM with the quantum system possessing an inherent pseudo-SUSY generated by the $Q$
operator (see \cite{Mos2} for an elaborate discussion on pseudo-SUSY).

 \section{\label {eg} Quasi-Hamiltonian $K$ and its classifications:}
 \numberwithin{equation}{section}

  A SSUSY model developed for $\mu=\lambda^2$ is of particular interest  for which the corresponding quasi-Hamiltonian $K$ is given by
\begin{eqnarray}
K &=& \left(\mathfrak{H}-\lambda\right)^2\nonumber\\
&=& \left(
      \begin{array}{cc}
        \left(\tha-\lambda\right)^2 & 0 \\
        0 & \left(\ttha-\lambda\right)^2 \\
      \end{array}
    \right).\label{k-1}
    \end{eqnarray}
 Also from (\ref{sup-ch}) and (\ref{quasi-h}) it follows
 \begin{equation}\label{k-2}
 K=\left(
     \begin{array}{cc}
       \mathcal{A}^+ \mathcal{A}^- & 0 \\
       0 & \mathcal{A}^- \mathcal{A}^+ \\
     \end{array}
   \right)
   \end{equation}
   Combining (\ref{k-1}) and (\ref{k-2}) gives
   \begin{equation}\label{A-pm}
   \mathcal{A}^+ \mathcal{A}^- = \left(\tha-\lambda\right)^2,\quad \mathcal{A}^- \mathcal{A}^+ = \left(\ttha-\lambda\right)^2.
   \end{equation}

   Let us factorize $\mathcal{A}^\pm$ in terms of a pair of general first-order differential quantities $\xi_1$ and $\xi_2$:
   \begin{equation}\label{Fact}
   \mathcal{A}^+ =\xi_1^\dag \xi_2^\dag, \quad \mathcal{A}^- =\xi_2 \xi_1
   \end{equation}
 where
 \begin{equation}\label{zeta}
 \xi_1 =\tilde{a}(x)\frac{d}{dx}+b_1(x),\quad \xi_2 =\tilde{a}(x)\frac{d}{dx}+b_2(x).
 \end{equation}
 Correspondingly
 \begin{eqnarray}
 \mathcal{A}^+ \mathcal{A}^- &=&  \xi_1^\dag \xi_2^\dag \xi_2 \xi_1 \label{a+a-}\\
 \mathcal{A}^- \mathcal{A}^+ &=&  \xi_2 \xi_1 \xi_1^\dag \xi_2^\dag \label{a-a+}
 \end{eqnarray}
 To be consistent with the perfect square form (\ref{A-pm}) we need to impose the compatibility constraint
 \begin{equation}
  \xi_2^\dag \xi_2 =\xi_1 \xi_1^\dag \label{constr}\\
  \end{equation}
 From (\ref{constr}) we are thus led to the relations
 \begin{equation}
b_2^2 -(\tilde{a} b_2)'= \tilde{a}(b_1-\tilde{a}')'+b_1(b_1-\tilde{a}'),
\end{equation}
 along with
 \begin{equation}\label{b12}
  \tilde{a} \tilde{a}''= \tilde{a}(b_1+ b_2)'-\tilde{a}'(b_1- b_2)+(b_1-b_2)(b_1+b_2).
 \end{equation}

 On using the constraint (\ref{constr}), we obtain from (\ref{susy-int}),(\ref{A-pm}),(\ref{a+a-}) and (\ref{a-a+})
 \begin{equation}\label{h1}
 \tha=\xi_1^\dag \xi_1 +\lambda =-\frac{d}{dx}\tilde{a}^2\frac{d}{dx}+ b_1^2-(\tilde{a}b_1)'+\lambda,
 \end{equation}
 Hence (\ref{V-0}) gives for $\tilde{V}^{(\alpha,\beta)}_{+}$
 \begin{equation}\label{V-1}
 \tilde{V}^{(\alpha,\beta)}_{+}=b_1^2-(\tilde{a}b_1)'+\lambda
 \end{equation}
In a similar way $\ttha$ is turns out to be
 \begin{equation}\label{h2}
 \ttha=\xi_2\xi_2^\dag +\lambda =-\frac{d}{dx}\tilde{a}^2\frac{d}{dx}+\tilde{a}(b_2-\tilde{a}')'+b_2(b_2-\tilde{a}')+\lambda
 \end{equation}
with the accompanying potential
\begin{equation}\label{V+1}
\tilde{V}^{(\alpha,\beta)}_{-}=\tilde{a}(b_2-\tilde{a}')'+b_2(b_2-\tilde{a}')+\lambda.
\end{equation}

 With the above relation at hand, we can speak of an intermediate Hamiltonian $\bha$ being superpartner to both $\tha$ and $\ttha$ i.e.
 \begin{equation}\label{h-12tld1}
 \tha=\xi_1^\dag \xi_1 +\lambda,\quad \bha=\xi_2^\dag\xi_2 +\lambda,\quad \ttha=\xi_2 \xi_2^\dag +\lambda.
 \end{equation}
  on using (\ref{constr}). Explicitly the intermediate Hamiltonian $\bha$ has the form
 \begin{equation}\label{h-tld}
 \bha=\xi_1 \xi_1^\dag +\lambda=\xi_2^\dag \xi_2 +\lambda=-\frac{d}{dx}\tilde{a}^2\frac{d}{dx}+\bar{V}^{(\alpha,\beta)},\quad \bar{V}^{(\alpha,\beta)}=b_2^2-(\tilde{a}b_2)'+\lambda.
 \end{equation}

 We therefore arrive at a triplet of Hamiltonians    $(\xi_1^\dag \xi_1, \xi_2^\dag\xi_2, \xi_2 \xi_2^\dag)$ of which the middle one plays a superpartner to the first and third components. In other words, we run into a position when the underlying equivalent Hermitian Hamiltonian of the Swanson model is determined from two standard SUSY Hamiltonians $(\xi_1^\dag \xi_1, \xi_2 \xi_2^\dag)$ and  $(\xi_2 \xi_2^\dag, \xi_2^\dag\xi_2)$. This is quite typical of a SSUSY formalism. However two things are new here. First is our employment of the Swanson model which goes beyond the standard Hermitian purview of quantum mechanics but has an equivalent Hermitian interpretation under a similarity transformation that is compatible with a pseudo-Hermitan theory. Second, rather than a canonical quantum condition of the harmonic oscillator we have been guided by a generalized form
 (\ref{gen-qnt-cond}). An off-shoot of such an implementation is that the component Hamiltonians  $\tha$, $\ttha$ and $\bha$ assume a PDM Schr\"odinger form \cite{bag1}.

  Let us note that the aforementioned feature of pairwise SUSY persists with other forms of the quasi-Hamiltonian such as $K=\mathfrak{H}^2-\frac{c^2}{4}$ or a more generalized form namely, $K=\mathfrak{H}^2-2\lambda \mathfrak{H}+\mu,\quad \lambda^2> \mu$.  In the first case the following type of factorization is implied
\begin{eqnarray}
   \mathcal{A}^+ \mathcal{A}^- &=& \left(\tha+\frac{c}{2}\right)\left(\tha-\frac{c}{2}\right),\label{A1-p}\\
   \mathcal{A}^- \mathcal{A}^+ &=& \left(\ttha-\frac{c}{2}\right)\left(\ttha+\frac{c}{2}\right)\label{A1-m}
   \end{eqnarray}
  under the constraint  $\xi_1 \xi_1^\dag +\frac{c}{2}=\xi_2^\dag \xi_2 -\frac{c}{2}$. We are thus led to the potentials
  \begin{eqnarray}
   &&\tha=-\frac{d}{dx}\tilde{a}^2\frac{d}{dx}+ b_1^2-(\tilde{a}b_1)'+\frac{c}{2},\quad \tilde{V}^{(\alpha,\beta)}_{+}(x)=b_1^2-(\tilde{a}b_1)'+\frac{c}{2},\label{triplet1}\\
   &&\ttha=-\frac{d}{dx}\tilde{a}^2\frac{d}{dx}+\tilde{V}^{(\alpha,\beta)}_{-}(x),\quad \tilde{V}^{(\alpha,\beta)}_{-}(x)=\tilde{a}(b_2-\tilde{a}')'+b_2(b_2-\tilde{a}')-\frac{c}{2},\label{triplet2}\\
   &&\bha=\xi_1 \xi_1^\dag +\frac{c}{2}=\xi_2^\dag \xi_2 -\frac{c}{2}=-\frac{d}{dx}\tilde{a}^2\frac{d}{dx}+\bar{V}^{(\alpha,\beta)},\quad \bar{V}^{(\alpha,\beta)}=b_2^2-(\tilde{a}b_2)'-\frac{c}{2}.\nonumber\\
   &&\label{triplet3}
   \end{eqnarray}

  On the other hand, in the second case, we have to go for the factorization
  \begin{eqnarray}
   \mathcal{A}^+ \mathcal{A}^- &=& \left(\tha-\lambda+\sqrt{\lambda^2-\mu}\right)\left(\tha-\lambda-\sqrt{\lambda^2-\mu}\right),\label{AG-p}\\
   \mathcal{A}^- \mathcal{A}^+ &=& \left(\ttha-\lambda-\sqrt{\lambda^2-\mu}\right)\left(\ttha-\lambda+\sqrt{\lambda^2-\mu}\right),\label{AG-m}
   \end{eqnarray}
   tied with the constraint $\xi_1 \xi_1^\dag +\sqrt{\lambda^2-\mu}=\xi_2^\dag \xi_2 -\sqrt{\lambda^2-\mu}$. These provide the following set of potentials
   \begin{eqnarray}
   &&\tha=-\frac{d}{dx}\tilde{a}^2\frac{d}{dx}+ \tilde{V}^{(\alpha,\beta)}_{+},\nonumber\\ &&\tilde{V}^{(\alpha,\beta)}_{+}(x)=b_1^2-(\tilde{a}b_1)'+\lambda+\sqrt{\lambda^2-\mu},\label{1111}\\&&\nonumber\\
   &&\ttha=-\frac{d}{dx}\tilde{a}^2\frac{d}{dx}+\tilde{V}^{(\alpha,\beta)}_{-},\nonumber\\ &&\tilde{V}^{(\alpha,\beta)}_{-}(x)=\tilde{a}(b_2-\tilde{a}')'+b_2(b_2-\tilde{a}')+\lambda-\sqrt{\lambda^2-\mu},\label{2222}\\&&\nonumber\\
   &&\bha=\xi_1 \xi_1^\dag +\lambda+\sqrt{\lambda^2-\mu}=\xi_2^\dag \xi_2+\lambda-\sqrt{\lambda^2-\mu}\nonumber\\ &&\qquad=\frac{d}{dx}\tilde{a}^2\frac{d}{dx}+\bar{V}^{(\alpha,\beta)},\bar{V}^{(\alpha,\beta)}=b_2^2-(\tilde{a}b_2)'+\lambda-\sqrt{\lambda^2-\mu}.\label{3333}
   \end{eqnarray}
   To sum up this section, we presented a set of classifications of the factorized quasi-Hamiltonian $K$ based on the factorization energies involved in the specific choice for $K$ that generates the corresponding factorization schemes. Such an analysis reminds us of the classification of the standard second-derivative SUSY counterpart spelt out explicitly in \cite{nw1}. Depending on the sign or vanishing of the guiding parameter, several possibilities have been discussed there much similar to the three possibilities for $K$ considered here. The main difference is that, in the present model the underlying structure of the Swanson Hamiltonian is non-Hermitian and to describe a pseudo-SUSY description for it in a second-derivative framework, we had to address its Hermitian equivalent partner that in principle can be transformed back to the non-Hermitian form (\ref{SW-H-expl}). On the other hand, the iterative higher-order SUSY approach of \cite{nw1} deals with a typically Hermitian formalism.

In the following we first study the case of an isotonic oscillator and then turns to the CPRS potential.

 \section{\label{eg}The case of an isotonic oscillator:}
 \numberwithin{equation}{section}

  Let us consider specifying for $a(x)$, $b(x)$, $b_1(x)$ and $b_2(x)$ the following choice :

  \begin{equation}\label{ex-1}
  a(x)= x^2
  \end{equation}

  \begin{equation}\label{ex-2}
  b(x)= \frac{1}{x} +\frac{cx}{x^2+d}
  \end{equation}

  \begin{equation}\label{ex-3}
 b_1(x)=\frac{c_1}{x}-c_2 x+\frac{c_3 x}{x^2+d}
 \end{equation}

 \begin{equation}\label{ex-4}
 b_2(x)=\frac{k_1}{x}+k_2 x+\frac{k_3 x}{x^2+d}
 \end{equation}
 where $c, d, c_1, c_2, c_3, k_1, k_2$ and $k_3$ are suitable constants.

Then from (\ref{rho}) we have for the spectral function $\ro$ the form
\begin{equation} \ro=\left[\frac{x^{\frac{c}{d}-1}}{(x^2+d)^\frac{c}{2d}}\right]^{-\frac{\alpha-\beta}{\tilde{\omega}}}e^{\frac{\alpha-\beta}{2\tilde{\omega}}\frac{1}{x^2}}.
\end{equation}
and for the potentials $\tilde{V}^{(\alpha,\beta)}_{+}(x)$, $\tilde{V}^{(\alpha,\beta)}_{-}(x)$ and $\bar{V}^{(\alpha,\beta)}(x)$ acquire the expressions
\begin{equation}
\tilde{V}^{(\alpha,\beta)}_{+}(x)=\frac{p}{x^2}+qx^2+ c\frac{rx^{2}+s}{(x^2+d)^2}+t\label{ex-1}
\end{equation}

\begin{eqnarray}
&&\tilde{V}^{(\alpha,\beta)}_{-}(x)\nonumber\\&=& \frac{c_1^2}{x^2}+c_2(c_2+3\sqrt{\tilde{\omega}})x^2+[\lambda+\sqrt{\tilde{\omega}}c_1^2-2c_1(c_2+\sqrt{\tilde{\omega}})]\nonumber\\
 &+& \frac{c_3(\sqrt{\tilde{\omega}}-2c_2)x^4+c_3(c_3-d\sqrt{\tilde{\omega}}+2c_1-2dc_2-2\sqrt{\tilde{\omega}})x^2+2dc_3(c_1-\sqrt{\tilde{\omega}})}{(x^2+d)^2}\label{ex-4}\\
&&\bar{V}^{(\alpha,\beta)}(x)\nonumber\\&=&\left[\frac{c_1}{x}-\frac{\sqrt{\tilde{\omega}}}{2}-\frac{c_3 x}{x^2+d}\right]^2- \sqrt{\tilde{\omega}}\left[-c_1+\frac{3\sqrt{\tilde{\omega}}}{2}x^2-\frac{3c_3x^2}{x^2+d}+\frac{2c_3 x^4}{(x^2+d)^2}\right]+\lambda\label{ex-5}
 \end{eqnarray}

where the various parameters stand for
\begin{eqnarray}
p&=&\frac{(\alpha-\beta)^2}{\tilde{\omega}}+\tilde{\omega}+2(\alpha+\beta), \quad q=\alpha+\beta+\frac{(\alpha-\beta)^2}{\tilde{\omega}}+2(\alpha+\beta),\nonumber\\ r&=&(2+c+2d)p-3d(\tilde{\omega}+\alpha+\beta) ,\quad s=2(1+d)p-d(\tilde{\omega}+\alpha+\beta),\nonumber\\ t&=&(c+\frac{3}{2})(\tilde{\omega}+\alpha+\beta)-2(c+1)p \\
c_1&=&\sqrt{p},\quad c_2=-\frac{3\sqrt{\tilde{\omega}}}{2},\quad c_3=c(1+\frac{1}{d})\sqrt{p}-\frac{c(\tilde{\omega}+\alpha+\beta)}{2\sqrt{p}}\\
k_1&=&-c_1,\quad k_2=2\sqrt{\tilde{\omega}}+c_2,\quad k_3=-c_3
\end{eqnarray}

 In all the three cases of the three potentials above, we find a typical combination of the harmonic oscillator, a centrifugal barrier and a non-polynomial term. In particular, for $r=s=0$, $\tilde{V}_+^{(\alpha,\beta)}$ depicts the isotonic oscillator while for $p=0$, it stands for the CPRS potential \cite{cprs}. However, if $p\neq0$, the CPRS potential is accompanied by a centrifugal barrier term.

 In the following section, we analyze the CPRS potential in some detail.

\section{\label{corr}The CPRS potential:}
\numberwithin{equation}{section}

We start with the conventional time-independent Schr\"odinger equation
\begin{equation}\label{cse-1}
\left[-\frac{d^2}{dy^2}+U(y)\right]\phi(y)=\epsilon \phi(y)
\end{equation}
under the influence of the potential $U$ and $\epsilon$ denotes the energy term. In terms of the superpotential $W(y)$ the potential can be interpreted as
\begin{equation}\label{cse-sup-pot-1}
U(y)=W^2(y)-\frac{dW(y)}{dy}+\varepsilon_0
\end{equation}
where $\varepsilon_0$ is the factorization energy.
\\
The change in variable
\begin{equation}\label{change-cse-1}
y=z(x)=\int^x \frac{dx'}{\tilde{a}(x')};
\end{equation}
transforms (\ref{cse-1}) into
\begin{equation}\label{cse-2}
\left[-\tilde{a}\frac{d}{dx}\tilde{a}\frac{d}{dx}+U\left(z(x)\right)\right]\chi(x)=\epsilon \chi(x)
\end{equation}
where $\chi(x)\equiv \phi(y(x))$.
The substitution $\chi(x)=\sqrt{\tilde{a}(x)}\psi(x)$ then leads us
\begin{equation}\label{cse-3}
\left[-\frac{d}{dx}\tilde{a}^2\frac{d}{dx}-\left(\frac{\tilde{a}\tilde{a}''}{2}+\frac{\tilde{a}'^2}{4}\right)+U\left(z(x)\right)\right]\psi(x)=\epsilon \psi(x).
\end{equation}
From (\ref{h-0}) we can thus identify $\tilde{V}^{(\alpha,\beta)}(x)$ as
\begin{equation}\label{cse-4}
\tilde{V}^{(\alpha,\beta)}_{+}(x)=-\left(\frac{\tilde{a}\tilde{a}''}{2}+\frac{\tilde{a}'^2}{4}\right)+ U\left(z(x)\right)
\end{equation}
implying that the Hamiltonian $\tHa$ shares the energy $E=\epsilon$.\

From equations (\ref{cse-4}),(\ref{cse-sup-pot-1}) and (\ref{h-0}) it follows that
\begin{eqnarray}
&& W^2\left(z(x)\right)-\tilde{a}(x)W'\left(z(x)\right)+\varepsilon_0 \nonumber\\
       &=&p_1 b\left(b-\frac{\tilde{a}'}{\sqrt{\tilde{\omega}}}\right)-q_1 \frac{\tilde{a}}{\sqrt{\tilde{\omega}}}b'+\frac{q_1}{2} \frac{\tilde{a}}{\sqrt{\tilde{\omega}}}\frac{\tilde{a}''}{\sqrt{\tilde{\omega}}}+\frac{p_1}{4}\left(\frac{\tilde{a}'}{\sqrt{\tilde{\omega}}}\right)^2
       +\frac{q_1}{2}\nonumber\\&&\label{cse-V-0}
\end{eqnarray}
where the various parameters stand for
\begin{eqnarray}
p_1 &=& \left[\frac{(\alpha-\beta)^2}{\tilde{\omega}}+\tilde{\omega}+2(\alpha+\beta)\right],\nonumber\\
q_1 &=& \left[\tilde{\omega}+(\alpha+\beta)\right].\label{par2}
\end{eqnarray}

Further comparing (\ref{cse-4}) with (\ref{V-1}) gives us an explicit form of $b_1$:
\begin{equation}\label{cse-b1}
b_1(x)=\frac{\tilde{a}'(x)}{2}+W\left(z(x)\right),\quad \lambda=\varepsilon_0.
\end{equation}

At this stage let us also mention that for any solution $\left(\epsilon,\phi(y)\right)$ of conventional Schr\"odinger equation (\ref{cse-1}) we can derive a corresponding solution $\left(\epsilon,\psi(x)\right)$ where \begin{equation}
\psi(x)=\frac{1}{\sqrt{\tilde{a}(x)}}\chi(x)=\frac{1}{\sqrt{\tilde{a}(x)}}\phi\left(z(x)\right)
\end{equation}

Generating an extended family for an exactly solvably potential starting from a seed potential is an interesting exercise within the realm of quantum mechanics. From a higher order SUSY point of view, explicit constructions of the partners of the simple harmonic potential have been made \cite{nw3} leading to, for example, Abraham-Moses potentials \cite{Fer1,nw4} and continuous families of anharmonic oscillator potential. We adopt a somewhat similar strategy to determine explicit forms of $\tilde{V}^{(\alpha,\beta)}_{+}(x)$ and $\bar{V}^{(\alpha,\beta)}(x)$ by making a suitable \textit{ansatz} for $\tilde{a}(x)$. In this regard, we now consider the following form of the CPRS potential \cite{cprs,fs}
\begin{equation}\label{cprs-pot}
U(y)=y^2+8\frac{2y^2-1}{(2y^2+1)^2}
\end{equation}
whose eigenvalues and wavefunctions are
\begin{eqnarray}
\epsilon_n&=&-3+2n,\qquad n=0,3,4,5,....\label{cprs-1}\\
\phi(y)&=&\frac{P_n(y)}{(2y^2+1)}e^{-\frac{y^2}{2}},\qquad n=0,3,4,5,....\label{cprs-2}
\end{eqnarray}
where $P_n(y)$ are related to Hermite polynomials by
\begin{equation}
P_n(y)=\left\{
           \begin{array}{ll}
             1, & \hbox{n=0;} \\
             H_n(y)+4nH_{n-2}(y)+4n(n-3)H_{n-4}(y), & \hbox{n=3,4,5,....}
           \end{array}
         \right.
\end{equation}
The corresponding superpotential is then obtained from (\ref{cse-sup-pot-1}):
\begin{equation}\label{cprs-3}
W(y)=y+\frac{4y}{2y^2+1},\qquad \varepsilon_0=-3.
\end{equation}
To proceed further we require a choice of $\tilde{a}(x)$ and we consider
\begin{equation}\label{cprs-4}
\tilde{a}(x)=\sqrt{\tilde{\omega}}x^\kappa, \quad 0\leq\kappa<1
\end{equation}
so that (\ref{change-cse-1}) gives
\begin{equation}\label{cprs-5}
z=\frac{1}{\sqrt{\tilde{\omega}}(1-\kappa)}x^{1-\kappa}
\end{equation}
for which
\begin{equation}
W(z(x))=\frac{1}{\sqrt{\tilde{\omega}}(1-\kappa)}x^{1-\kappa}+2\sqrt{\tilde{\omega}}(1-\kappa)\frac{1}{x^{1-\kappa}}.
\end{equation}

Correspondingly $b(x)$ can be determined following the equation (\ref{cse-V-0}):
\begin{equation}\label{cprs7}
b(x)=\frac{1}{\sqrt{\tilde{\omega}}(1-\kappa)}x^{1-\kappa}+\frac{2\sqrt{\tilde{\omega}}(1-\kappa)+\frac{\kappa}{2}}{x^{1-\kappa}}+\varrho(x)
\end{equation}
where $\varrho(x)$ satisfies the equation
\begin{equation}\label{varrho}
\varrho^2(x)-\varrho'(x)+2\left[\frac{1}{\sqrt{\tilde{\omega}}(1-\kappa)}x^{1-\kappa}+2\sqrt{\tilde{\omega}}(1-\kappa)\frac{1}{x^{1-\kappa}}\right]\varrho(x)
+(3+\frac{\sqrt{\tilde{\omega}}}{2})=0
\end{equation}
and $p_1,q_1$ present in (\ref{cse-V-0}) can be found:
$$p_1=1,\quad q_1=\sqrt{\tilde{\omega}}.$$

The parameters $\alpha,\beta,\tilde{\omega}$ are then determined from (\ref{par2}):
$$\alpha=-\beta,\quad \tilde{\omega}=\frac{1}{2}[1+\sqrt{1-16\alpha^2}],\quad \alpha^2\leq 16.$$

Also from (\ref{cse-b1}) we get
\begin{equation}\label{cprs-7}
b_1(x)=\frac{1}{\sqrt{\tilde{\omega}}(1-\kappa)}x^{1-\kappa}+\frac{\sqrt{\tilde{\omega}}(4-3\kappa)}{2}\frac{1}{x^{1-\kappa}}
\end{equation}
so that (\ref{b12}) gives
\begin{equation}\label{cprs-8}
b_2(x)=-\frac{1}{\sqrt{\tilde{\omega}}(1-\kappa)}x^{1-\kappa}+\frac{\sqrt{\tilde{\omega}}(4-\kappa)}{2}\frac{1}{x^{1-\kappa}}.
\end{equation}
Then from (\ref{V-1}),(\ref{V+1}) and (\ref{h-tld}), after some simple calculations we obtain
\begin{eqnarray}
\tilde{V}^{(\alpha,\beta)}_{+}(x) &=& \tilde{V}^{(\alpha,\beta)}_{-}(x)\nonumber\\ &=&\frac{\kappa(2-3\kappa)\tilde{\omega}}{4}\frac{1}{x^{2(1-\kappa)}}+\frac{1}{\tilde{\omega}(1-\kappa)^2}x^{2(1-\kappa)}
+ 4\tilde{\omega}(1-\kappa)^2\frac{x^{2(1-\kappa)}-\frac{\tilde{\omega}(1-\kappa)^2}{2}}{\left[x^{2(1-\kappa)}+\frac{\tilde{\omega}(1-\kappa)^2}{2}\right]^2},
\nonumber\\
&&\label{cprs-9}\\&&\nonumber\\
\bar{V}^{(\alpha,\beta)}(x) &=& \frac{\tilde{\omega}(2-\kappa)(4-5\kappa)}{4}\frac{1}{x^{2(1-\kappa)}}+\frac{1}{\tilde{\omega}(1-\kappa)^2}x^{2(1-\kappa)}+2.\label{cprs-10}
\end{eqnarray}

Furthermore, for $\kappa=0,\mbox{ and } \tilde{\omega}=1$, we get $\tilde{a}=1$ and then our result for $\tilde{V}^{(\alpha,\beta)}_{+}(x)$ represents CPRS potential whereas $\bar{V}^{(\alpha,\beta)}(x)=\frac{2}{x^2}+x^2+2$ represents isotonic oscillator.

\section{Conclusion}
\numberwithin{equation}{section}

In this paper a new general approach is suggested towards the construction of second-derivative schemes for the Swanson oscillator both in the SUSY as well as in pseudo-SUSY scenarios. The Swanson model deals with an extended harmonic oscillator Hamiltonian when terms involving squares of annihilation and creation operators are present with different strengths of coupling constants. As a result the natural Hermiticity of the ordinary harmonic oscillator is lost. However, Swanson Hamiltonian is both $\mathcal{PT}$- symmetric and pseudo-Hermitian for a suitable choice of the metric. In this article we have applied the criterion of a generalized quantum condition first to write down the corresponding Hamiltonian form and then interpret it from a SUSY description in a second-derivative framework. In the resulting SSUSY scheme we have determined the corresponding set of a triplet of component Hamiltonians  and discussed a concrete scheme concerning the isotonic oscillator. We have then formulated a pseudo-SUSY framework by defining a pair of suitable pseudo-adjoint operators. We discussed various test cases. Another interesting conclusion of our work is that by means of a coordinate transformation we can transform the  Schr\"odinger equation to address the CPRS potential and determine the component Hamiltonians that are relevant to the second-derivative framework.

\noindent{\Large{\textbf{Acknowledgement}}}:\\
One of us (P.M.) thanks the Council of Scientific and Industrial Research, New Delhi for the award of Senior Research Fellowship. We are grateful to the referee for his/her valuable suggestions.

}

\newpage
\begin{thebibliography}{99}
\bibitem{Ben1}C.M.Bender and S.Boettcher, \textit{Phys.Rev.Lett.} \textbf{24} 5243 1998.
\bibitem{Ben2}C.M.Bender, \textit{Rep.Prog.in Phys.} \textbf{70}, 947 2007.
\bibitem{Mos1}A.Mostafazadeh, \textit{Int.J.Geom.Meth.Mod.Phys.} \textbf{7}, 1191 2010.
\bibitem{Mos2}A.Mostafazadeh, \textit{J.Math.Phys.} \textbf{43}, 205 2002.
\bibitem{Mos3}A.Mostafazadeh, \textit{J.Math.Phys.} \textbf{43}, 2814 2002.
\bibitem{Mos4}A.Mostafazadeh, \textit{J.Math.Phys.} \textbf{43}, 3944 2002.
\bibitem{Fer}F.M.Fern\'{a}ndez, J.Garcia, I. Semoradora and M.Znojil, \textit{Ad hoc Hilbert space in Quantum Mechanics} arXiv:1405.7284.
\bibitem{Swan}M.S.Swanson, \textit{J.Math.Phys. } \textbf{45}, 585 2004.
\bibitem{Jones}H.F.Jones, \textit{J.Phys.A:Math.Gen.} \textbf{38},1 741 2007.
\bibitem{Mus}D.P.Musumbu, H.B.Geyer and W.Heiss, \textit{J.Phys.A:Math.Gen.} \textbf{40}, F75 2007.
\bibitem{Qes1}C.Quesne, \textit{J.Phys.A:Math.Gen.} \textbf{40}, F745 2007
\bibitem{Asi}P.E.G. Assis and A.Fring, \textit{J.Phys.A:Math.Theor.} \textbf{42}, 015203 2009
\bibitem{Qes2}C.Quesne, \textit{J.Phys.A:Math.Gen.} \textbf{41}, 244022 2008
\bibitem{Yes1}O.Yesiltas, \textit{Phys. Scr. } \textbf{87} 045013 2013      .
\bibitem{BT}B.Bagchi and T.Tanaka, \textit{Phys.Lett.A } \textbf{372}, 5390 2008.
\bibitem{Fri1}B.Bagchi and A.Fring, \textit{Phys.Lett.A } \textbf{373}, 4307 2009.
\bibitem{Fri2}S.Dey, A.Fring and B.Khantoul, \textit{J.Phys.A:Math.Theor.} \textbf{46}, 335304 2013.
\bibitem{Rr}R.Roychoudhury, B.Roy and P.P.Dube, \textit{J.Math.Phys. } \textbf{54}, 012104 2013.
\bibitem{CDMT}O.Cherbal, M.Drir, M.Maamache and D.A.Trifonov, \textit{SIGMA } \textbf{6}, 2010 096
\bibitem{DG}Q. Duret and F. Gieres Non-Hermitian Hamiltonians and supersymmetric quantum mechanics  Preprint 2004.
\bibitem{Gra}E.-M. Grafe, H.J.Korsch, A. Rush and R. Schubert, \textit{Classical and quantum dynamics in the (non-Hermitian) Swanson oscillator} arXiv:1409.6456.
\bibitem{bag1}B.Bagchi, C.Quesne and R.Roychoudhury, \textit{J.Phys.A:Math.Gen. } \textbf{38}, L647 2005.
\bibitem{yes}O.Yesiltas, \textit{J.Phys.A } \textbf{44}, 305305 2014.
\bibitem{And1}A.Andrianov, N.V.Borisov and M.V.Ioffe, \textit{Phys.Lett.A } \textbf{105}, 19 1984.
\bibitem{And2}A.Andrianov, N.V.Borisov, M.I.Eides and M.V.Ioffe, \textit{Phys.Lett.A} \textbf{109}, 143 1985.
\bibitem{And3}A.Andrianov, M.V.Ioffe and V.Spiridonov, \textit{Phys.Lett.A } \textbf{174}, 273 1993.
\bibitem{Fer1}D.J.Fern\'{a}ndez C., \textit{Int.J.Mod.Phys.A } \textbf{12}, 171 1997.
\bibitem{Fer2}D.J.Fern\'{a}ndez C. and V.Hussin, \textit{J.Phys.A:Math.Gen.} \textbf{32}, 3603 1999.
\bibitem{bag2}B.Bagchi, S.Mallik and C.Quesne, \textit{Int.J.Mod.Phys.A } \textbf{17},5 1 2002.
\bibitem{Can}F.Cannata and M.V.Ioffe, \textit{Phys.Lett.B} \textbf{278}, 399 1992.
\bibitem{Pup}A.M. Pupasov, B.F.Samsonov and U.Gunther \textit{J.Phys.A:Math.Theor.} \textbf{40}, 10557 2007.
\bibitem{Andr}A.A.Andrianov, M.V.Ioffe and D.N.Nishnianidze, \textit{Theor.Math.Phys.A} \textbf{104}, 1129 1995.
\bibitem{pin}B.Bagchi, P.Gorain,C.Quesne and R.Roychoudhury, \textit{Mod.Phys.Lett.A } \textbf{19}, 2765 2004.
\bibitem{nw1}D.J.Fern\'{a}ndez C. and F.Fern\'{a}ndez-Garcia \textit{AIP Conf.Proc. } \textbf{744}, 236 2005.
\bibitem{nw3}D.J.Fern\'{a}ndez C., V.Hussin and B.Mielnik \textit{Phys.Lett.A } \textbf{244}, 309 1998.
\bibitem{nw4}D.J.Fern\'{a}ndez C., M.L.Glasser and L.M.Nieto \textit{Phys.Lett.A } \textbf{240}, 15 1998.
\bibitem{cprs}J.F.Cari$\tilde{\mbox{n}}$ena, A.M.Perelomov, M.F.Ra$\tilde{\mbox{n}}$ada and M.Santander, \textit{J.Phys.A:Math.Gen.} \textbf{41}, 085301 2008.
\bibitem{fs}J.M.Fellows and R.A.Smith, \textit{J.Phys.A:Math.Theor.} \textbf{42}, 335303 2009.
\end{thebibliography}
\end{document}